\documentclass[11pt,epsf,letterpaper]{article}%
\usepackage{color}
\usepackage{amsmath}
\usepackage{amsfonts}
\usepackage{verbatim}
\usepackage{amssymb}
\usepackage{graphicx}
\usepackage{epstopdf}
\usepackage{mathrsfs}
\usepackage{epsfig}%
\providecommand{\U}[1]{\protect\rule{.1in}{.1in}}
\begin{document}

\title{Hairy black holes: stability under odd-parity perturbations and existence of
slowly rotating solutions}
\author{Andr\'{e}s Anabal\'{o}n$^{(1)}$, Ji\v{r}\'{\i} Bi\v{c}\'{a}k$^{(2)}$ and Joel
Saavedra$^{(3)}$\\\textit{$^{(1)}$Departamento de Ciencias, Facultad de Artes Liberales y}\\\textit{Facultad de Ingenier\'{\i}a y Ciencias, Universidad Adolfo
Ib\'{a}\~{n}ez,}\\\textit{Av. Padre Hurtado 750, Vi\~{n}a del Mar, Chile}\\\textit{$^{(2)}$ Institute of Theoretical Physics, Charles University, V
Hole\v{s}ovi\v{c}k\'{a}ch 2,}\\\textit{180 00 Prague 8, Czech-Republic}\\\textit{$^{(3)}$Instituto de F\'\i sica, Pontificia Universidad Cat\'olica de
Valpara\'\i so,} \\\textit{ Casilla 4059, Valpara\'{\i}so, Chile} \\andres.anabalon@uai.cl, jiri.bicak@mff.cuni.cz, joel.saavedra@ucv.cl}
\maketitle

\begin{abstract}
We show that, independently of the scalar field potential and of specific
asymptotic properties of the spacetime (asymptotically flat, de Sitter or
anti-de Sitter), any static, spherically symmetric or planar, black hole or
soliton solution of the Einstein theory minimally coupled to a real scalar
field with a general potential is mode stable under linear odd-parity
perturbations. To this end, we generalize the Regge-Wheeler equation for a
generic self-interacting scalar field, and show that the potential of the
relevant Schr\"{o}dinger operator can be mapped, by the so-called
S-deformation, to a semi-positively defined potential. With these results at
hand we study the existence of slowly rotating configurations. The frame
dragging effect is compared with the Kerr black hole.

\end{abstract}

\section{Introduction}

In the light of discoveries of last decades like cosmic acceleration,
associated with \textquotedblleft dark energy", or more definitive indications
of the existence of \textquotedblleft dark matter", Einstein's classical
general relativity has occurred again \textquotedblleft on the firing line",
to use Cliff Will's metaphor from the October 1972 issue of Physics Today.
Numerous modifications of Einstein's theory assume the existence of scalar
field(s). Scalar fields are part of the inflationary paradigm and appear
conspicuously in string theory and supergravity. The reality of a fundamental
scalar field appears to be supported by the discovery of the
Brout-Englert-Higgs boson. The interest on the interaction of scalar fields
and gravity is however older than inflation and supergravity and can be traced
back to the well known Brans-Dicke theory and the black hole no-hair theorems.
The original, and well known, no-hair theorem of Bekenstein states that a
convex potential is incompatible with the existence of a static black hole in
asymptotically flat spacetime \cite{Bekenstein:1995un}. This is also supported
by the fact that massive scalar fields develop a power-law singularity at the
horizon when solved in the Kerr-Newman background \cite{Bicak:1978pn}.
Bekenstein's theorem was generalized and we understand now that the relevant
requirement for an asymptotically flat black hole to exist in the presence of
a regular, non-trivial, scalar field profile is the existence of a negative
region of the scalar field potential \cite{Heusler:1992ss, Sudarsky:1995zg}.

An interest has been revived in black holes in scalar-tensor theories and
their perturbations; for solutions with a real minimally coupled scalar field,
see \cite{Anabalon:2012tu}; for solutions in theories with non-minimal
derivative coupling, see \cite{Hordenski}; for stability, see
\cite{Harper:2003wt}. In particular, a Japanese group studied black hole
perturbations in a general gravitational theory with Lagrangian given by an
arbitrary function of the Ricci scalar and the Chern-Simons pseudoinvariant
\cite{MotoSuy1, MotoSuy2}. Most recently, the authors associated with this
group tackled a technically involved problem of perturbations of black holes
in the most general scalar-tensor theory in which all field equations are of
the second order. Their Lagrangian includes, for example, the Brans-Dicke
theory, $f(R)$ gravity, the non-minimal coupling to the Gauss-Bonnet term,
etc. The odd-parity perturbations have been tackled first \cite{Kobay1}; most
recently, the even-parity sector was also analyzed \cite{Kobay2}. The authors
concentrate primarily on deriving perturbation equations from the second-order
actions. In both cases they present general conditions, necessary but not
sufficient, for the (gradient) stability of a static, spherically symmetric solution.

We remain in the framework of classical four dimensional general relativity
and consider perturbations and stability of static, spherically symmetric,
planar and hyperbolic hairy black holes in asymptotically flat or
asymptotically (anti-) de Sitter spacetimes in which a scalar field is
minimally coupled to gravity (for a recent review, see \cite{AA}). Then we
analyze their odd-parity perturbations following the general treatment of the
\textquotedblleft axial\textquotedblright\ perturbations of spherically
symmetric (not necessarily vacuum) spacetimes by Chandrasekhar \cite{Chandra}.
We prove the mode stability with respect to general perturbations in the
odd-parity sector.

Very recently, Bhattacharya and Maeda \cite{BhaMa} arrived at a conclusion,
based on the slow-rotation approximation, that
Bocharova-Bronnikov-Melnikov-Bekenstein black hole \cite{BBM, Bek} with a
scalar field coupled conformally to gravity cannot rotate (for related
discussions se also \cite{Astorino:2013sfa, Hassaine:2013cma, Bardoux:2013swa}%
). We provide examples when the rotation of a general class of hairy black
holes is admissible. Although in the odd-parity case, the gravitational
perturbations are fully decoupled from the scalar field perturbations, they
can strongly be influenced by the background scalar field. The situation
resembles general perturbations of charged (Reissner-Nordstr\"{o}m) black
holes where the background electric field influences gravitational
perturbations (cf. \cite{Mo, JiBi, BiDv}). The right-hand side of Einstein's
equations is given in terms of the energy-momentum tensor of (possibly strong)
background scalar field and odd-parity perturbations of the metric. As a
consequence, the metric component like $g_{\varphi t}$, which implies the
dragging of inertial frames outside a rotating black hole, depends on the
character of the background scalar field. We illustrate this effect.

Rotating black holes and boson stars are known to exist when a complex scalar
field is minimally coupled to Einstein theory \cite{Herdeiro:2014goa}.
However, much less is known when the scalar field is real. To construct the
slowly rotating solutions it is necessary to specify the background. Thus, we
use the static black hole family originally found in \cite{Anabalon:2012ta}.
This family of solutions is the most general four dimensional hairy black hole
family with a single real scalar field and contains all other exact black
holes available in the literature. The details can be found in \cite{AA}.

The outline of this article is as follows. In the second section we present
the proof that for minimally coupled scalar fields with arbitrary
self-interaction, the spectrum of the generalized Regge-Wheeler equation is
always positive. The proof is done in detail for spherical geometries and
generalized to planar trasnversal geometries. The third section introduces the
perturbative frame-dragging computation for the hairy black hole. We briefly
describe the hairy black hole geometries before comparing the frame draging
effect with the Kerr black hole. Finally we present a discussion of our main
results. Our conventions are such that the Riemann tensor and the Ricci scalar
of a sphere are positive in an orthonormal basis. The metric signature is
$(-,+,+,+)$ and we set $\kappa=8\pi G$, $c=1$.

\section{Odd-parity perturbations and generalized Regge-Wheeler equation}

Here we follow closely \cite{Chandra}. We shall consider a minimally coupled
real scalar field with an arbitrary potential, $V(\phi)$. The field equations
are%
\begin{align}
E_{\mu\nu}  &  \equiv G_{\mu\nu}-\kappa T_{\mu\nu}=0\text{ },\label{EFEQ}\\
T_{\mu\nu}  &  =\partial_{\mu}\phi\partial_{\nu}\phi-g_{\mu\nu}\left[
\frac{1}{2}\left(  \partial\phi\right)  ^{2}+V(\phi)\right]  \ , \label{2}%
\end{align}
where $G_{\mu\nu}$ is the Einstein tensor and a possibly non-vanishing
consmological constant can be included in $V(\phi)$. The perturbed metric
reads%
\begin{equation}
ds^{2}=-A(r)dt^{2}+B(r)dr^{2}+C(r)\left[  \frac{dz^{2}}{\left(  1-kz^{2}%
\right)  }+\left(  1-kz^{2}\right)  \left(  d\varphi+k_{1}dt+k_{2}%
dr+k_{3}dz\right)  ^{2}\right]  \ ,
\end{equation}
where $k_{1}$, $k_{2}$ and $k_{3}$ are functions of ($t,r,z$). $A(r),$ $B(r)$
and $C(r)$ are the metric functions parameterizing the most general static
background solution of a scalar-tensor theory. For asymptotically locally AdS
solutions $k=\pm1\,\ $or $0$ \cite{Lemos:1994xp}. Asymptotically flat or de
Sitter solutions have $k=1$. The scalar field is taken to be of the form%

\begin{equation}
\phi=\phi_{0}(r)+\epsilon\Phi\left(  t,r,z\right)  \ , \label{pert}%
\end{equation}
where $\phi_{0}$ is the background field. The metric perturbations $\left(
k_{1},k_{2},k_{3}\right)  $ are all taken to be first order in $\epsilon$.
Since any surface of constant $(t,r)$ is of constant curvature, we consider
only axisymmetric perturbations, without any loss of generality (for more
details of the spherically symmetric case, see \cite{Chandra}). The Einstein
field equations are truncated at first order in $\epsilon$. This yields the
vanishing of $\Phi$. Indeed, using the notation introduced in equation
(\ref{EFEQ}) and the zeroth-order (background)\ equations we find that%

\begin{align}
E_{r}^{t}  &  =\epsilon\kappa\frac{d\phi_{0}}{dr}\frac{\partial_{t}\Phi}%
{A(r)}+O(\epsilon^{2})=0\text{ },\label{oddscalar1}\\
E_{z}^{r}  &  =-\epsilon\kappa\frac{d\phi_{0}}{dr}\frac{\partial_{z}\Phi
}{B(r)}+O(\epsilon^{2})=0\text{ },\\
E_{t}^{t}  &  =\frac{\epsilon\kappa}{B(r)}\left(  \frac{d\phi_{0}}{dr}%
\partial_{r}\Phi+BV_{1}\Phi\right)  +O(\epsilon^{2})=0\text{ },
\label{oddscalar3}%
\end{align}
where $V_{1}$ arises from the expansion of the scalar field potential around
the background configuration,
\begin{equation}
V_{1}=\left.  \frac{dV}{d\phi}\right\vert _{\phi=\phi_{0}}\text{ }.
\end{equation}
Equations (\ref{oddscalar1})-(\ref{oddscalar3}) imply that $\partial_{t}%
\Phi=0=\partial_{z}\Phi$ and $\frac{d\phi_{0}}{dr}\partial_{r}\Phi=-BV_{1}%
\Phi$. This information simplifies equation for $E_{r}^{r}:$%

\begin{equation}
E_{r}^{r}=2\epsilon\kappa V_{1}\Phi+O(\epsilon^{2})=0\text{ }.
\end{equation}
Thus, it follows that $\Phi=0$. Using the zeroth-order equations, it is
possible to check that the remaining equations are satisfied up to linear
order in $\epsilon$ if the following system of equations is satisfied:%
\begin{align}
\frac{\partial}{\partial r}\left[  C\sqrt{\frac{A}{B}}\left(  \partial
_{z}k_{2}-\partial_{r}k_{3}\right)  \right]  +\frac{\partial}{\partial
t}\left[  C\sqrt{\frac{B}{A}}\left(  \partial_{t}k_{3}-\partial_{z}%
k_{1}\right)  \right]   &  =0\text{ },\label{A1}\\
\frac{\partial}{\partial z}\left[  \frac{A}{C}(1-kz^{2})^{2}\left(
\partial_{z}k_{2}-\partial_{r}k_{3}\right)  \right]  +\frac{\partial}{\partial
t}\left[  (1-kz^{2})\left(  \partial_{r}k_{1}-\partial_{t}k_{2}\right)
\right]   &  =0\text{ },\label{A2}\\
\frac{\partial}{\partial z}\left[  C\sqrt{\frac{B}{A}}(1-kz^{2})^{2}\left(
\partial_{z}k_{1}-\partial_{t}k_{3}\right)  \right]  +\frac{\partial}{\partial
r}\left[  (1-kz^{2})\frac{C^{2}}{\sqrt{AB}}\left(  \partial_{r}k_{1}%
-\partial_{t}k_{2}\right)  \right]   &  =0\text{ }. \label{A3}%
\end{align}
Introducing the variable $Q=CA^{1/2}B^{-1/2}(1-kz^{2})^{2}\left(  \partial
_{z}k_{2}-\partial_{x}k_{3}\right)  $, equations (\ref{A1})-(\ref{A2}) yield%
\begin{align}
\frac{A^{1/2}}{CB^{1/2}\left(  1-kz^{2}\right)  ^{2}}\frac{\partial
Q}{\partial r}  &  =-\partial_{t}^{2}k_{3}+\partial_{t}\partial_{z}k_{1}\text{
},\label{B1}\\
\frac{\sqrt{AB}}{C^{2}}\frac{1}{\left(  1-kz^{2}\right)  }\frac{\partial
Q}{\partial z}  &  =-\partial_{t}\partial_{r}k_{1}+\partial_{t}^{2}k_{2}\text{
}. \label{B2}%
\end{align}
The combination $\partial_{r}$(\ref{B1})+$\partial_{z}$(\ref{B2}) can be
written in terms of $Q,$%
\begin{equation}
\frac{C^{2}}{\sqrt{AB}}\frac{\partial}{\partial r}\left[  \frac{A^{1/2}%
}{CB^{1/2}}\frac{\partial Q}{\partial r}\right]  +(1-kz^{2})^{2}\frac
{\partial}{\partial z}\left[  \frac{1}{\left(  1-kz^{2}\right)  }%
\frac{\partial Q}{\partial z}\right]  =\frac{C}{A}\partial_{t}^{2}Q\text{ }.
\end{equation}
This equation can be solved by separation of variables. Writting
$Q=q(r,t)D(z),$ we obtain:%
\begin{align}
\frac{C^{2}}{\sqrt{AB}}\frac{\partial}{\partial r}\left[  \frac{A^{1/2}%
}{CB^{1/2}}\frac{\partial Q}{\partial r}\right]  -\lambda Q  &  =\frac{C}%
{A}\partial_{t}^{2}Q\text{ },\label{eq}\\
\left(  1-kz^{2}\right)  ^{2}\frac{\partial}{\partial z}\left[  \frac
{1}{\left(  1-kz^{2}\right)  }\frac{\partial D}{\partial z}\right]   &
=-\lambda D\text{ }. \label{gegenbauer}%
\end{align}

Let us now concentrate on the case with $k=1$; we shall briefly comment on the
other cases later. Setting $z=\cos\theta$ in equation (\ref{gegenbauer})
allows to identify $C_{l+2}^{-3/2}(\theta)=D(z)$ with a Gegenbauer polynomial
with $\lambda=\left(  l-1\right)  \left(  l+2\right)  $ where $l\geq1$
holds\footnote{That $l\geq1$ follows from the definition of the Gegenbauer
polynomials in terms of the Legendre polynomials: $C_{l+2}^{-3/2}(\theta
)=\sin^{3}\theta\frac{d}{d\theta}\frac{1}{\sin\theta}\frac{dP_{l}(\theta
)}{d\theta}$.}. The master variable in this case is $\Psi(r^{\ast
},t)=q(r,t)C^{-1/2}$ where $\frac{\partial}{\partial r}=\frac{B^{1/2}}%
{A^{1/2}}\frac{\partial}{\partial r^{\ast}}$. Inserting all this information
in equation (\ref{eq}) yields the master equation%

\begin{equation}
\frac{\partial^{2}\Psi}{\partial r^{\ast2}}+\left(  \frac{1}{2C}\frac{d^{2}%
C}{dr^{\ast2}}-\frac{3}{4C^{2}}\left(  \frac{dC}{dr^{\ast}}\right)
^{2}-\lambda\frac{A}{C}\right)  \Psi=\partial_{t}^{2}\Psi\text{ }.
\end{equation}
The mode stability can be studied using the Fourier decomposition of the
master variable, $\Psi=\int\Psi_{\omega}e^{i\omega t}dt$, which yields%

\begin{equation}
\mathcal{H}\Psi_{\omega}\equiv-\frac{d^{2}\Psi_{\omega}}{dr^{\ast2}}+\left(
\lambda\frac{A}{C}+\frac{3}{4C^{2}}\left(  \frac{dC}{dr^{\ast}}\right)
^{2}-\frac{1}{2C}\frac{d^{2}C}{dr^{\ast2}}\right)  \Psi_{\omega}=\omega
^{2}\Psi_{\omega}\text{ }. \label{RW}%
\end{equation}
The scalar field perturbation vanishes, however equation (\ref{RW}) depends on
the background scalar field through its influence on the background metric. In
vacuum, $A=1-2m/r$, $C=r^{2}$, and equation (\ref{RW}) becomes the
Regge-Wheeler equation. The operator $\mathcal{H}$ is not manifestly positive,
however, its spectrum is positively defined as follows from\footnote{For a
more detailed discussion, see \cite{Ishibashi:2003ap}.}%

\begin{equation}
\int dr^{\ast}\left(  \Psi_{\omega}\right)  ^{\ast}\mathcal{H}\Psi_{\omega
}=\int dr^{\ast}\left[  \left\vert D\Psi_{\omega}\right\vert ^{2}%
+V_{S}\left\vert \Psi_{\omega}\right\vert ^{2}\right]  -\left.  \left(
\Psi_{\omega}D\Psi_{\omega}\right)  \right\vert _{Boundary}\text{ },
\label{RW1}%
\end{equation}
where $D=\frac{d}{dr^{\ast}}+S$ and%
\begin{equation}
V_{S}=\lambda\frac{A}{C}+\frac{3}{4C^{2}}\left(  \frac{dC}{dr^{\ast}}\right)
^{2}-\frac{1}{2C}\frac{d^{2}C}{dr^{\ast2}}+\frac{dS}{dr^{\ast}}-S^{2}\text{ }.
\end{equation}
Choosing $S=\frac{1}{2C}\frac{dC}{dr^{\ast}},$ we find%
\begin{equation}
V_{S}=\lambda\frac{A}{C}\text{ }. \label{RW2}%
\end{equation}
Therefore $l\geq1\Longrightarrow\lambda\geq0\Longrightarrow V_{S}\geq0$
whenever $A>0$, namely in any static region of the spacetime. From equations
(\ref{RW1}-\ref{RW2}) it follows that all the spherically symmetric four
dimensional hairy configurations are mode stable under odd-parity
perturbations. To reach this conclusion it is necessary that%

\begin{equation}
\left.  \left(  \Psi_{\omega}D\Psi_{\omega}\right)  \right\vert _{Boundary}%
=0\text{ },
\end{equation}
which requires that the perturbation vanishes at the horizon. This is not a
very strong requirement, as follow from \cite{Kay:1987ax}; linear stability
under the boundary condition $\Psi=0$ at the horizon implies stability under
boundary conditions with $\Psi$ taking a finite value at the horizon.

Note that $k\,$\ goes into the perturbation equation (\ref{RW}) only through
$\lambda\,$. When $k=0$ the requirement that the perturbations are everywhere
well defined is satisfied only if $\lambda>0$ which implies $V_{S}\geq0$. The
equation for the angular part is just the equation for the harmonic oscillator
with frequency $\sqrt{\lambda}$. When $k=-1,$ further analysis is required.

\section{Slowly rotating hairy black holes}

Here we want to establish the existence of slowly rotating hairy black holes.
To this end we shall consider only stationary perturbations with $k_{2}%
=k_{3}=0$ and $k_{1}=\omega(r)$. In this case equation (\ref{A3}) yields%

\begin{equation}
\omega=-c_{1}\int\frac{\sqrt{AB}}{C^{2}}dr+c_{2}\text{ }, \label{omega}%
\end{equation}
where $c_{1}$ and $c_{2}$ are two integration constants. To warm up, let us
consider now the case of the Schwarzschild black hole. We have $\sqrt{AB}=1$
and $C=r^{2},$ so it follows that%

\begin{equation}
\omega=\frac{c_{1}}{3r^{3}}+c_{2}\text{ }.
\end{equation}
Hence, choosing $c_{2}=0$ and $c_{1}=3Ma,$ we find that the perturbed metric
is the Schwarzschild metric plus the perturbation $g_{t\varphi}=\frac
{Ma(1-z^{2})}{r}$ which, when terms proportional to $a^{2}$ are neglected,
coincides exactly with the Kerr metric in the Boyer-Lindquist coordinates.

Now let us consider the hairy black hole family \cite{AA, Anabalon:2012ta}.
The following configurations are exact background solutions of the Einstein
equations\footnote{For a single real scalar field, the scalar field equation
is a consequence of the Einstein equations through the conservation of the
energy-momentum tensor.} (\ref{EFEQ}):%

\begin{equation}
ds^{2}=\Omega(x)\left[  -f(x)dt^{2}+\frac{\eta^{2}dx^{2}}{f(x)}+\frac{dz^{2}%
}{1-kz^{2}}+\left(  1-kz^{2}\right)  d\varphi^{2}\right]  \text{ ,} \label{S1}%
\end{equation}

\begin{equation}
\Omega(x)=\frac{\nu^{2}x^{\nu-1}}{\eta^{2}\left(  x^{\nu}-1\right)  ^{2}%
}\text{ }, \label{O}%
\end{equation}

\begin{equation}
f(x)=\frac{x^{2-\nu}\left(  x^{\nu}-1\right)  ^{2}\eta^{2}k}{\nu^{2}}+\left(
\frac{1}{\nu^{2}-4}-\frac{x^{2}}{\nu^{2}}\left(  1+\frac{x^{-\nu}}{\nu
-2}-\frac{x^{\nu}}{\nu+2}\right)  \right)  \alpha+\frac{1}{l^{2}}\text{ ,}
\label{S2}%
\end{equation}
with energy-momentum tensor given by equation (\ref{2}), scalar field
potential and background scalar field by:%

\begin{align}
V(\phi)  &  =\frac{\alpha}{\kappa\nu^{2}}\left[  \frac{\nu-1}{\nu+2}\sinh(\phi
l_{\nu}\left(  \nu+1\right)  )-\frac{\nu+1}{\nu-2}\sinh(\phi l_{\nu}\left(
\nu-1\right)  )+4\frac{\nu^{2}-1}{\nu^{2}-4}\sinh(\phi l_{\nu})\right]
\nonumber\\
&  -\frac{\left(  \nu^{2}-4\right)  }{2\kappa l^{2}\nu^{2}}\left[  \frac
{\nu-1}{\nu+2}\exp(-\phi l_{\nu}\left(  \nu+1\right)  )+\frac{\nu+1}{\nu
-2}\exp(\phi l_{\nu}\left(  \nu-1\right)  )+4\frac{\nu^{2}-1}{\nu^{2}-4}%
\exp(-\phi l_{\nu})\right]  \text{ ,} \label{W}%
\end{align}
and%

\begin{equation}
\phi_{0}=l_{\nu}^{-1}\ln x\text{ ,}%
\end{equation}
where parameter $\eta$ is the unique integration constant; it arises in a
non-standard form which allows to write the solution in terms of a
dimensionless radial coordinate $x$ and $l_{\nu}^{-1}=\sqrt{\frac{\nu^{2}%
-1}{2\kappa}\text{. }}$The metric and the potential are invariant under the
change $\nu\rightarrow-\nu$, therefore, it is possible to take $\nu\geq1$.
Furthermore, it should be noted that the asymptotic region is at $x=1$ which
can be seen from the pole of order two in $\Omega(x)$. There are two
solutions, depending on whether the scalar field is negative, $x\in(0,1),$ or
positive, $x\in(1,\infty),$ which allows to cover all the values of the scalar
field potential. The metric is regular for any value of $x\neq0$ and
$x\neq\infty$ as can be seen from the introduction of advanced and retarded
coordinates, $u_{\pm}=t\mp\int\frac{\eta}{f(x)}dx$. The scalar field and the
geometries are singular at $x=0$ and $x=\infty$ but these singularities can be
covered by event horizons. The metric reduces to the Schwarzschild-(A)dS
solution in Schwarzschild-Droste coordinates when $\nu=1$ and setting
$x=1\pm1/\left(  \eta r\right)  $. The mass of the spherically symmetric
solution, computed with the Hamiltonian method \cite{Hertog:2004dr}, yields%
\begin{equation}
M=\pm\frac{\alpha+3\eta^{2}}{6\eta^{3}G}\text{ .}%
\end{equation}
where the $\pm$ depends on whether one is considering the branch where $x>1$
or $x<1$.

In analogy with the Kerr solution, the slowly rotating hairy black hole is a
deformation of the static one plus $g_{t\varphi}=\omega_{\nu}(1-z^{2}%
)\Omega(x).$ The metric component $g_{t\varphi}$ determines the frame dragging
potential (see e.g. \cite{Misner:1974qy} Ex. 3.4). We find that%

\begin{equation}
\omega_{\nu}=\bar{c}_{1}\frac{x^{2-\nu}}{\nu^{2}(\nu^{2}-4)}(\left(
\nu-2\right)  x^{2\nu}+\left(  4-\nu^{2}\right)  x^{\nu}-2-\nu)+\bar{c}%
_{2}\text{ ;}%
\end{equation}
requiring that $\omega_{\nu}(x=1)=0$ fixes $\omega$ up to an overall
multiplicative constant,%

\begin{equation}
\omega_{\nu}=\bar{c}_{1}\left(  \frac{x^{2-\nu}}{\nu^{2}(\nu^{2}-4)}(\left(
\nu-2\right)  x^{2\nu}+\left(  4-\nu^{2}\right)  x^{\nu}-2-\nu)+\frac{1}%
{\nu^{2}-4}\right)  \text{ .}%
\end{equation}

To measure the deviation from the slowly rotating Kerr solution we plot the
ratio $\omega/\omega_{\nu=1}$ versus the square root of the areal function
$\sqrt{\Omega(x)}$. The integration constant $\eta$ has units of inverse of
length squared. Hence the $x$-axis is measured in units of $\eta^{-1}$ (e.g.
km, parsec, etc.). In Figure 1 it can be seen that there is a smooth departure
from the Kerr frame dragging as both coincide when $\nu$ approaches 1 as well
as asymptotically for large $\sqrt{\Omega(x)}$. It should be noticed that the
departure from Kerr can be important and that the horizon can be located at
any point in the graph. Indeed, the location of the horizon is defined by the
equation $f(x_{+})=0$, which has solution for any $x_{+}$ by adjusting the
value of $\alpha$ in expression (\ref{S2}).

\begin{figure}[ptb]
\epsfig{file=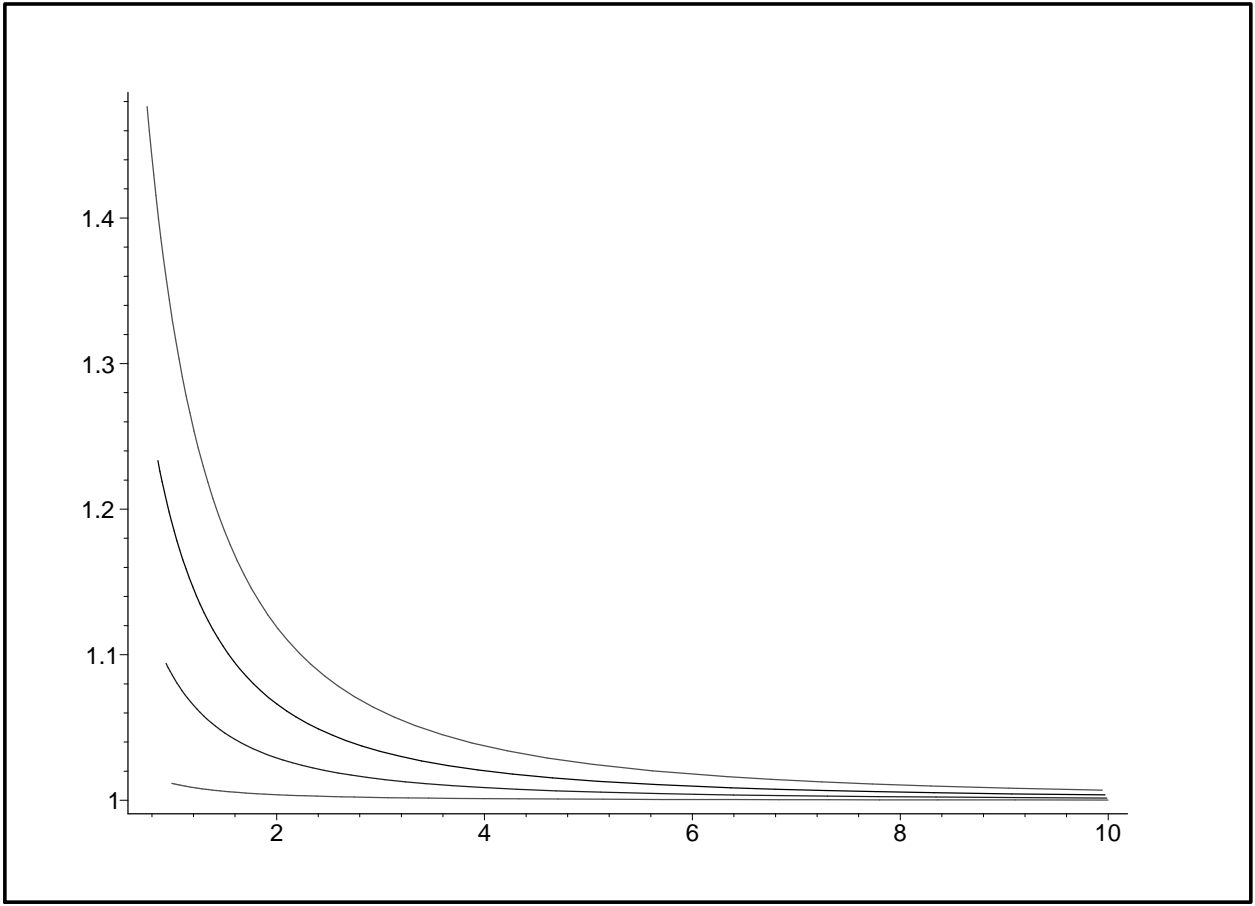,width=12cm,angle=270}\caption{The ratio
$\omega/\omega_{\nu=1}$ versus the square root of the areal function,
$\sqrt{\Omega(x)}$, for different values of $\nu$. The plots are for
$\nu=1.2,\nu=2.1,\nu=3$ and $\nu=4$ (from down up).}%
\end{figure}

\section{Conclusions}

In this paper we addressed the issue of odd-mode stability in a rather general
class of scalar-tensor theories. We proved that for a minimally-coupled real
scalar field, independently of its self-interaction and of the asymptotic
properties of the spacetime (asymptotically flat, de Sitter or anti-de
Sitter), any static soliton or black hole solution is mode stable under these
perturbations. The situation is such that the scalar field only contributes
through the backreaction of the background solution and the dynamics is
dictated by the linearized Einstein equations. This is in contrast of what
happens with the spherically symmetric mode where the only propagating mode is
the scalar one \cite{Harper:2003wt}. The linearized Einstein-scalar field
equations allowed us to study the existence of slowly rotating hairy black
hole solutions. Using the hairy black hole family \cite{AA, Anabalon:2012ta},
we have shown that there is no obstruction for the existence of rotating hairy
black holes and that they can have a behavior that strongly departs from the
Kerr solution. Indeed, it would be very interesting to study the even modes
outside the spherically symmetric regime, namely when the scalar and the
tensor perturbations interact non-trivially. We leave this question open to
further research.

\section{Acknowledgments}

Research of A.A. is supported in part by the Fondecyt Grants N%
${{}^o}$
11121187, 1141073. Research of J.S. is supported in part by the Fondecyt
Grants N%
${{}^o}$
1110076, 1110230. J.B. acknowledges the support from CONICYT and is grateful
for a wonderful hospitality offered by colleagues at Ponitificia Universidad
Catolica de Valparaiso, Universidad Adolfo Iba\~{n}ez de Vi\~{n}a del Mar and
Universidad Andres Bello de Santiago. J.B. acknowledges the support from the
Czech Science Foundation, grant N%
${{}^o}$
14-37086G (Albert Einstein Center).

\end{document}